\documentclass{ws-procs10x7}

\def \gtsim    {\relax\ifmmode{\mathrel{\mathpalette\oversim >}}
                  \else{$\mathrel{\mathpalette\oversim >}$}\fi}
\def \ltsim    {\relax\ifmmode{\mathrel{\mathpalette\oversim <}}
                  \else{$\mathrel{\mathpalette\oversim <}$}\fi}
\def\oversim#1#2{\lower4pt\vbox{\baselineskip0pt \lineskip1.5pt
            \ialign{$\mathsurround=0pt#1\hfil##\hfil$\crcr#2\crcr\sim\crcr}}}
\newcommand{\invfb}{\mbox{${\rm fb}^{-1}$}}

\newcommand{\degr  }{\mbox{$^{\circ}$}}
\newcommand{\pt}  {\mbox{$p_{T}$}}

\newcommand{\menergy} {\mbox{${E\!\!\!\!/}$}}
\newcommand{\mpt}{\mbox{${p\!\!\!/_T}$}}
\newcommand{\meff}  {\mbox{$M_{{\rm eff}}$}}
\newcommand{\qqbar}{\mbox{$q\overline{q}$}}

\newcommand{\epem} {\mbox{$e^+e^-$}}

\newcommand{\tauh} {\mbox{$\tau_{{\rm h}}$ } }
\newcommand{ \stauone}  {\mbox{$\tilde{\tau}_{1}$}}
\newcommand{ \stauonep} {\mbox{$\tilde{\tau}_{1}^{+}$}}
\newcommand{ \stauonem} {\mbox{$\tilde{\tau}_{1}^{-}$}}

\newcommand{ \staup}    {\mbox{$\tilde{\tau}^{+}$}}
\newcommand{ \staum}    {\mbox{$\tilde{\tau}^{-}$}}

\newcommand{ \schionezero }{\mbox{$\tilde{\chi}_{1}^{0}$}}
\newcommand{ \schitwozero }{\mbox{$\tilde{\chi}_{2}^{0}$}}

\newcommand{ \schionepm }{\mbox{$\tilde{\chi}_{1}^{\pm}$}}

\def \etal     {\relax\ifmmode{et \; al.}\else{$et \; al.$}\fi}

\begin{document}

\title{Minimal SUGRA Model and Collider Signals}

\author{ R. Arnowitt,{$^\dagger$}
Bhaskar Dutta,{$^*$} T. Kamon{$^\dagger$} and V. Khotilovich{$^\dagger$}}

\address{$^\dagger$Department of Physics, Texas A$\&$M University, 
College Station, TX 77807, USA\\
$^*$Department of Physics, University of Regina, 
Regina, Saskatchewan S4S 0A2, Canada }

\twocolumn[\maketitle
\abstract{
The SUSY signals
in the dominant stau-neutralino coannihilation region 
at a 500(800) GeV linear collider are investigated. 
The region is consistent with the WMAP measurement of
the cold dark matter relic density as well as
all other current experimental bounds within the mSUGRA framework.
The signals are characterized
by an existence of very low-energy tau leptons in the final state
due to small mass difference between
$\stauone$ and $\schionezero$ (5-15 GeV).
We study the accuracy of the mass difference measurement
with a 1\degr\ active mask to reduce
a huge SM two-photon background.
}]

\section{Introduction}

The recent measurement of cold dark matter (CDM) relic density
 from WMAP\cite{sp} along with 
the Higgs mass bound and the $b\rightarrow s \gamma$ constraint have restricted the
parameter space significantly\cite{dark} within the framework of
minimal supergravity (mSUGRA) model.\cite{sugra1,sugra2}
One prominent parameter space is the region where
the mass difference ($\Delta M$) between
the lighter stau ($\stauone$) and the
lightest neutralino ($\schionezero$) is  about 5-15 GeV.
This small mass difference allowed the $\stauone$ to coannihilate in the early
universe along with the $\schionezero$
in order to produce the current amount of the CDM ($\schionezero$).
The coannihilation region has a large extension for $m_{1/2}$   
up to  1-1.5 TeV,
and can be explored at the LHC.
The main difficulty, however,  
in probing this region is to detect very low-energy taus 
in the final state of the SUSY events due to
the small $\Delta M$ value.  
 
In this paper, we report a feasibility study of measuring the small mass
difference in this $\stauone$-$\schionezero$ coannihilation region 
at a 500 GeV linear collider (LC).

\section{mSUGRA Parameter Space}

The  mSUGRA model depends on only four  parameters and one sign. 
These are $m_0$ (the universal soft breaking mass at the 
GUT scale $M_G$); $m_{1/2}$ (the universal gaugino soft breaking mass at $M_G$); 
$A_0$ (the universal cubic soft breaking mass at $M_G$); 
$\tan\beta = \langle H_2 \rangle / \langle H_1 \rangle$ 
at the  electroweak scale; and the sign of $\mu$, the Higgs mixing 
parameter in the superpotential ($W_{\mu} = \mu H_1 H_2$).  

\begin{figure}
\vspace*{-0.8cm}
\epsfxsize12.9pc
\figurebox{16pc}{32pc}{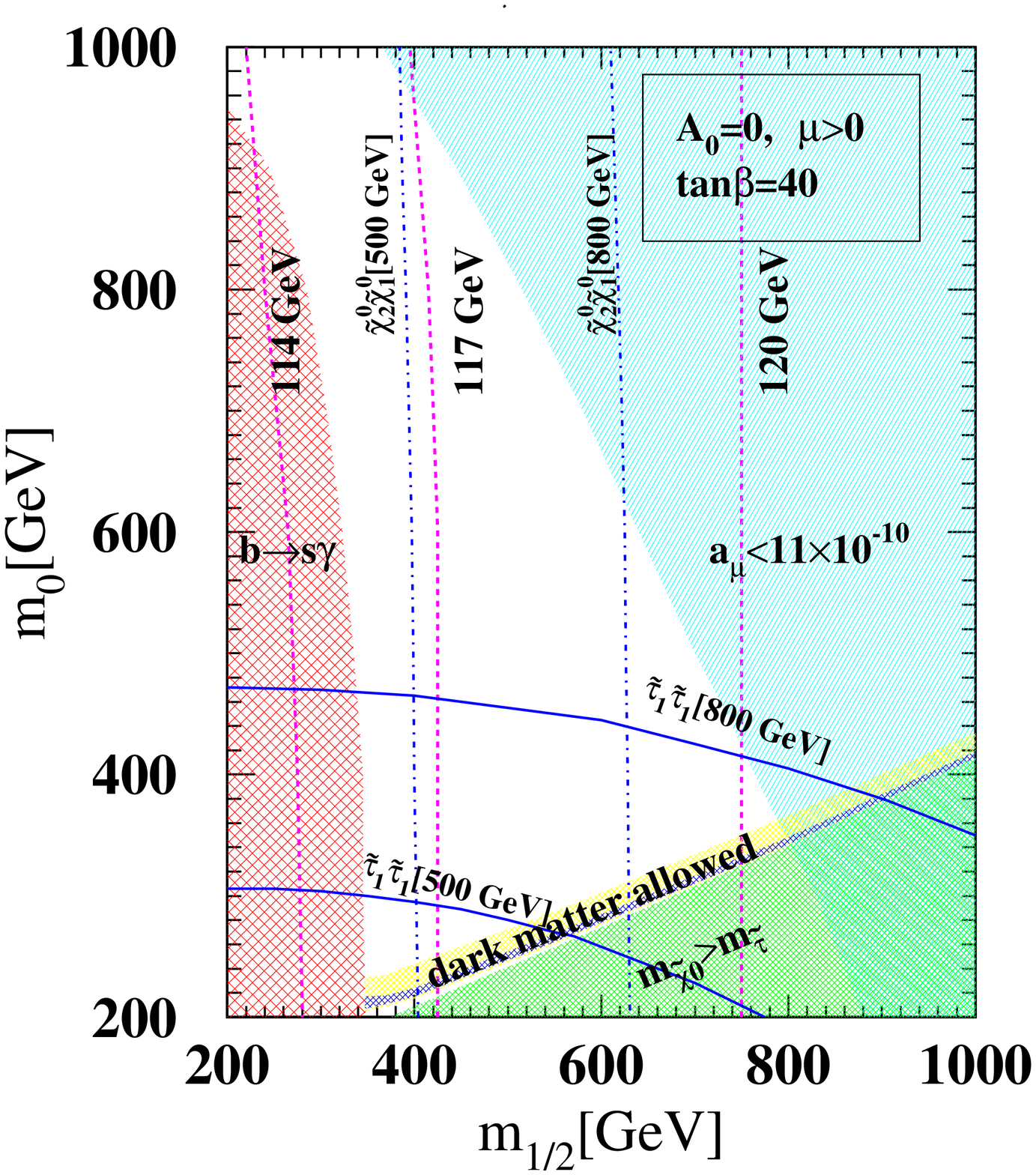}
\caption{Allowed region in the $m_0$-$m_{1/2}$ plane from the relic density 
constraint  for $\tan\beta$ =  40, 
$A_0  = 0$ and $\mu >0$. The details are provided in text.} 
\label{WMAP_allowed_region}
\end{figure}

Figure~\ref{WMAP_allowed_region} is an example of the allowed region
in the $m_0$-$m_{1/2}$ plane for $\tan\beta$ =  40  with $A_0$ = 0 and  $\mu > 0$.
The most important experimental
results  
for limiting the parameter space are:
(1)~The light Higgs mass bound (pink lines in the figure) from LEP:\cite{higgs1} 
$M_h > 114$ GeV;
(2)~The $b\rightarrow s + \gamma$ branching ratio (brick red region):\cite{bsgamma}
 $2\times10^{-4} < {\cal B}(B \rightarrow X_s \gamma) <
4.5\times10^{-4}$;
(3)~Previous CDM ($\schionezero$) density bounds of 
$0.07 < \Omega_{\schionezero} h^2 < 0.21 $ (yellow band)
from balloon flights (Boomerang, Maxima, 
Dasi, etc.) and
the 2$\sigma$ bound of $0.095 < \Omega_{\schionezero} h^2 <0.129$
(blue band) 
from WMAP\cite{sp};
(4)~The bound on the lightest chargino mass:\cite{aleph}
$\schionepm >$ 104 GeV;
(5)~Possible muon magnetic moment anomaly (light blue region
to be excluded if $\delta a_\mu > 11\times10^{-10}$).\cite{BNL}

It is striking to learn that only two SUSY production processes
can be studied at a 500 GeV LC: $\stauonep\stauonem$ and 
$\schitwozero\schionezero$.
The kinematical reaches via the 
$\stauonep\stauonem$ and $\schitwozero\schionezero$ production
are also shown in Fig.~\ref{WMAP_allowed_region}.
The maximum reach in $m_{1/2}$ along the coannihilation band
can be expected via 
$\epem \rightarrow  \stauonep \stauonem \rightarrow 
( \tau^+ \schionezero) + ( \tau^- \schionezero)$.

We use the hadronic final state of tau ($\tauh$)
since it has  larger branching ratios. 
Due to the small $\Delta M$ value, 
the taus in the final states are low energy 
and hence harder to detect.

\section{SUSY Signals at 500 GeV LC}

In order to optimize the event selection cuts,
we choose three points of $m_0$ = 205, 210 and 220 GeV
for $m_{1/2}$ = 360 GeV, $\tan\beta = 40$, $\mu > 0$, and $A_0 = 0$.
The SUSY masses given by {\tt ISAJET}\cite{isajet} are summarized in
Table~\ref{table:mSUGRA}.
There are two major Standard Model (SM)  background processes: 
(i)~four-fermion final state $\bar\nu \nu \tau^+ 
\tau^-$ arising from processes such as diboson ($WW$, $ZZ$)
production,  and 
(ii)~two-photon 
processes $e^+ e^- \rightarrow \gamma^* \gamma^* +  e^+ e^-\rightarrow \tau^+ \tau^-$ 
(or $q \bar{q}$) + $e^+ e^-$ 
where the final state $e^+ e^-$ pair are at a  small angle to the beam pipe and 
the $\qqbar$ jets fake a $\tau^+ \tau^-$ pair. 

The production cross-sections for SUSY ({\tt ISAJET})
and SM four-fermion processes 
({\tt WPHACT}\cite{wphact}) are listed in
Table~\ref{table:Xsec} 
for a 500 GeV LC. 
We choose with right handed (RH) polarized
electron beams to enhance 
the $\stauonep \stauonem$ events over  
the $\schionezero \schitwozero$  and SM four-fermion events.  

\begin{table}
\caption{Masses (in GeV) of SUSY particles in three representative scenarios of 
$\Delta M$
for $m_{1/2}$ = 360 GeV, $\tan\beta = 40$, $\mu > 0$, and $A_0 = 0$. }
\label{table:mSUGRA}
\begin{center}
\begin{tabular}{|l c | c c c c |}
\hline 
 MC & $m_0$ & $M_{\schitwozero}$ &  $M_{\stauone}$ & 
$M_{\schionezero}$ &
        $\Delta M$  \\
 Pt. & & & & &  \\
\hline  \hline
1 & 205 &   274.2 &  147.2 &  142.5 &  4.7   \\
2 & 210 & 274.2 &  152.0 &  142.5 &  9.5  \\
3 & 220 & 274.3 &  161.6 &  142.6 &  19.0 \\
\hline 
\end{tabular}
\end{center}
\end{table}

\begin{table}
\caption{SUSY and SM production cross sections 
($\sigma \cdot B(\tau \rightarrow \tauh)^{2}$ in fb)  for 
polarization for electron beams of ${\cal P}(e^-) = -0.9$(RH).}
\label{table:Xsec}
\begin{center} 
\begin{tabular}{|l c | c| }
\hline 
SUSY Pt. 1. & $\schionezero\schitwozero$ & 0.43  \\
	& $\stauonep\stauonem$ & 28.25 \\
	\hline
SUSY Pt. 2. & $\schionezero\schitwozero$ & 0.39 \\
	& $\stauonep\stauonem$ & 25.85  \\
\hline
SUSY Pt. 3. & $\schionezero\schitwozero$ & 0.38  \\
	& $\stauonep\stauonem$ & 22.95   \\
\hline 
\hline 
\multicolumn{2}{| l |}{SM (four fermion process)}    & 7.84  \\
\hline
\end{tabular}
\end{center}
\end{table}

\begin{table*}
\caption{Event selection criteria for  the RH (${\cal P} = -0.9$) case.}
\label{table:EventSelectionCuts}
\begin{center}
\begin{tabular}{ |l | c| }
\hline 
Variable(s) & Cuts \\
\hline \hline  
$N_{jet}$($E_{jet}>3$ GeV)  & 2 \\
$\tau_h$ ID  & 1, 3 tracks\\
\hline
Jet acceptance &  $|\cos(\theta_{jet})| < 0.65$\\
 &  $-0.6 < \cos[\theta(j_2,p_{vis})] < 0.6$\\\hline
Missing \pt (\mpt) & $>$ 5 GeV \\ \hline
Acoplanarity  & $> 40\degr$ \\
\hline
Veto on EM clusters & No EM cluster in $5.8\degr < \theta < 28\degr$ with $E > 2$ GeV\\ 
 or electrons & No electrons within $\theta > 28\degr$ with $\pt > 1.5$ GeV \\
 \hline
Beam mask (1\degr (or 2\degr) - 5.8\degr)  &
 No EM cluster with $E>100$ GeV \\
\hline
\end{tabular}
\end{center}
\end{table*}

In Table~\ref{table:EventSelectionCuts}, 
we summarize the event selection criteria for the  RH case.
The Monte Carlo (MC) events 
are generated, simulated and analyzed using the following programs: 
{\tt ISAJET}\cite{isajet} to generate SUSY events; 
{\tt WPHACT}\cite{wphact} for SM backgrounds;
{\tt TAUOLA}\cite{TAUOLA} for tau decay; 
a LC detector simulation\cite{dark} to reconstruct
jets with JADE algorithm.\cite{JADE}\
In our calculation, beamstrahlung and bremsstrahlung are
included in both {\tt ISAJET} and {\tt WPHACT}. 

The accepted number of signal and background events   
are summarized
in Tables~\ref{table:Nevent_500invfb} and~\ref{table:Nevent_500invfbsm}.  
It should be noted that the number of SM $\gamma\gamma$ events
with the forward electrons just below 3\degr\ are 11400.
The acceptances for $\stauonep\stauonem$ events are
11.2\%, 5.9\%, and 0.86\% for $\Delta M$= 19, 9.5, and
4.7 GeV, respectively with 1\degr mask. 
The acceptance
drops fast as $\Delta M$ goes below 5 GeV. 
For example, 0.23\% for  $\Delta M$= 3.8 GeV ($m_0$ = 204 GeV).
We see the robust discovery significance for 
the signal events for 
$\Delta M \gtsim$ 5 GeV with 1\degr\ mask
in Table~\ref{table:LC500_Results}.
We conclude that the mask is 
essential to detect SUSY events in this region of parameter space. 

\begin{table}
\caption{Number of SUSY events expected with 500 fb$^{-1}$ for
the RH case.}
\label{table:Nevent_500invfb}
\begin{center}
\begin{tabular}{|l| rrr|}
\hline
Process & $\Delta M$ =   4.7  & 9.5 & 19  \\
\hline \hline
$\schitwozero \schionezero$
        &    15 & 26 & 29 \\
\hline
$\staup \staum$
        &    122 & 786 & 1283\\
\hline
\end{tabular}
\end{center}
\end{table}

\begin{table}
\caption{Number of SM events expected with 500 fb$^{-1}$. }
\label{table:Nevent_500invfbsm}
\begin{center}
\begin{tabular}{|l l | c |}
\hline
\multicolumn{2}{| l |}{SM four-fermion} & 129 \\
\hline
SM $\gamma\gamma$ & 2-5.8\degr\ Mask   & 248 \\
 & 1-5.8\degr\ Mask   & 2 \\
\hline
\end{tabular}
\end{center}
\end{table}

\section{Measurement of Stau Neutralino Mass Difference}

Since  $\Delta M$ is small, it needs to be measured with a very good accuracy. 
We choose  the  invariant mass  $\meff\
\equiv M(j_1,j_2,\menergy)$ of two $\tau$-jets
and missing energy as a key discriminator.
We  generate high statistics MC samples for the SM and various SUSY events 
(by changing
the $m_0$ value) and prepare the templates of the \meff\ distributions
for the SM, $\schionezero\schitwozero$, and $\stauonep\stauonem$ events.

\begin{figure}
\epsfxsize16.0pc
\figurebox{16pc}{32pc}{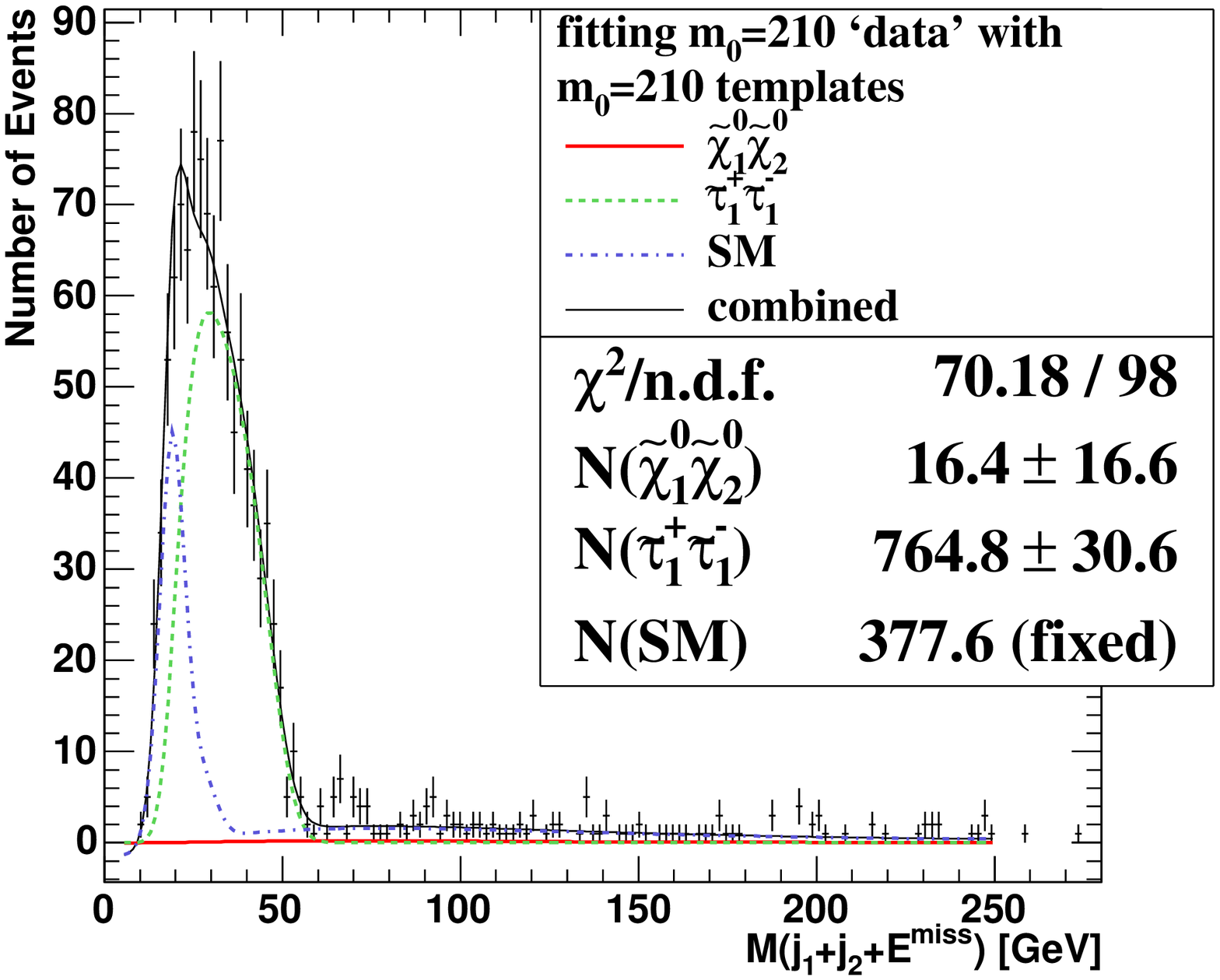}
\caption{$\meff\ (\equiv M(j_1,j_2,\menergy))$ distributions
for a 500 \invfb\ MC samples for SUSY ($m_0$ = 210 GeV) and SM events,
being fitted to the templates for $m_0$ = 210 GeV.}
\label{fits210211}
\end{figure}

We then generate the MC samples equivalent to 500 \invfb\ of luminosity
for particular $\Delta M$ values and
fit them with the template functions. 
For example, in Fig.~\ref{fits210211}
we show the fitting of the  500 \invfb\ MC samples for Point 2 with
the templates for $m_0$ = 210 GeV and calculate 
the $\chi^2$ of the fits.  
Here the $\chi^2$ value is calculated as 
$\chi^2=\sum_{i} \biggl(
{ { N_i - \sum_{j} C_{j}F_i^{j} }\over{ \sigma_i } } \biggr) ^2
$ where $N_i$ is the number of events in $i$-th \meff\ bin
of the 500 \invfb\ sample, 
$C_j F_i^j$ is the corresponding value for the template  ``$j$''
where $j$ is for 
SM, $\stauonep\stauonem$ or  $\schionezero\schitwozero$ 
processes.
$C_j$ is a normalization parameter and a free variable
except for the SM process.
This is because we should
be able to measure the SM events very well before
we discover SUSY events.

\begin{table}
\caption{Significance ($N_S/ {\sqrt{N_B}}$) with 500 \invfb\ for 
SUSY discovery using 1\degr\ mask.}
\label{table:LC500_Results}
\begin{center}
\begin{tabular}{|c|  c c c|}
\hline   
Process (RH) & $\Delta M$= 4.7 & 9.5 & 19 \\
\hline \hline
$\staup\staum$ & 10 & 63 & 101 \\\hline
\end{tabular}
\end{center}
\end{table}

\begin{figure}
\epsfxsize14pc
\figurebox{16pc}{32pc}{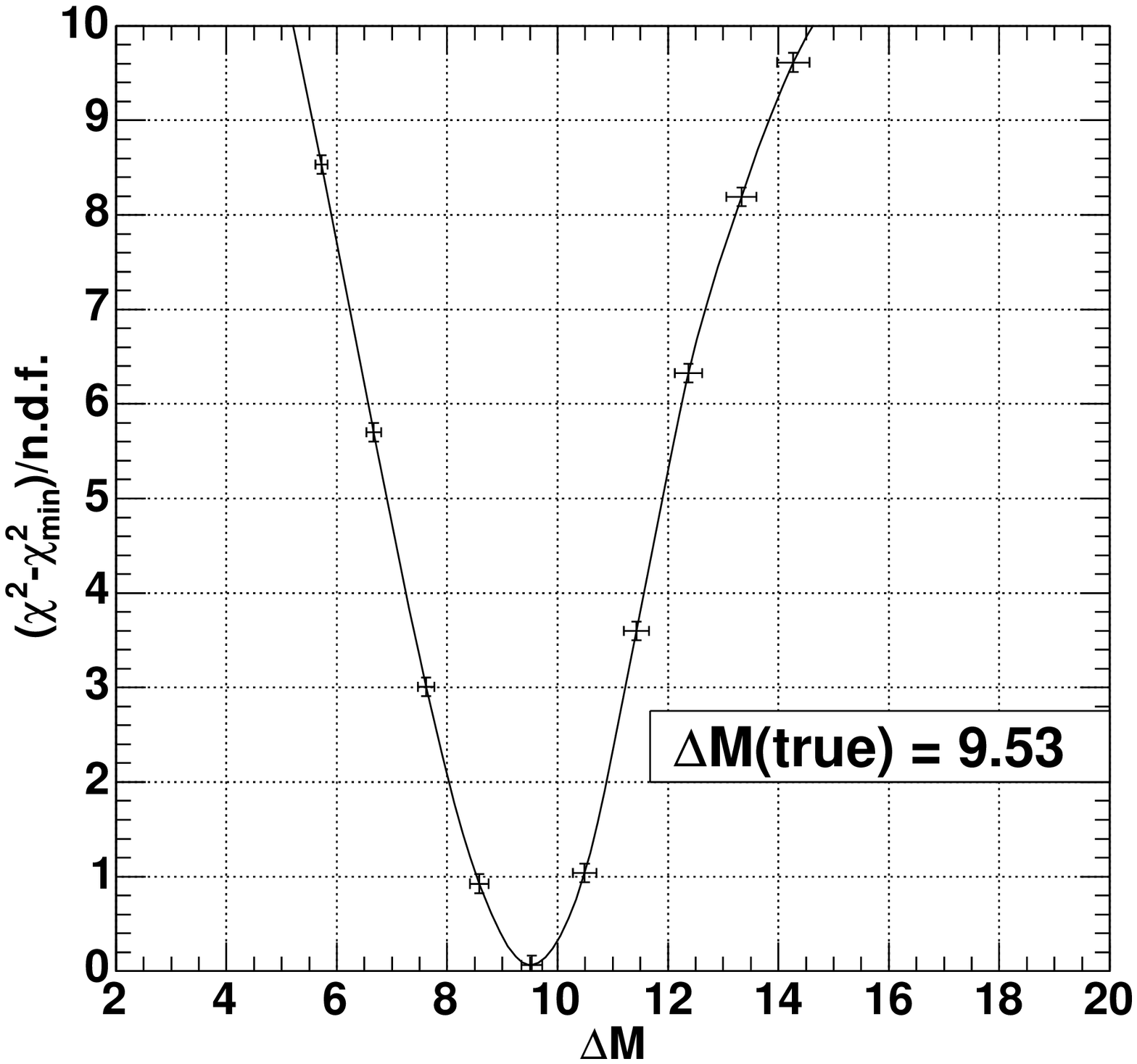}
\caption{$\chi^2$ (defined in the text) 
for Point 2 as a function of $\Delta M$.}
\label{chisq}
\end{figure}

We scan the range of $m_0$ = 203-220 GeV
and plot the $\Delta \chi^2 \equiv \chi^2 - \chi^2_{{\rm min}}$ 
in Fig.~\ref{chisq}.
The $\Delta \chi^2$ value is minimum for the template for $m_0$ = 210 GeV.
We find that 1$\sigma$ in the $\Delta \chi^2$ corresponds to $9.5\pm 1$ GeV,
where the true value of $\Delta M$ for the Point 2 is 9.53 GeV.

We repeat the same study for different stau masses i.e. 
for different $\Delta M$ values
and two different beam mask designs (1\degr\ and 2\degr).
For $\Delta M \sim$ 5 GeV, 
a beam mask of $1^\circ$ is crucial. 
The accuracy of mass determination for
is summarized
in Table~\ref{table:dM_accuracy},
showing 
the uncertainties are at a level of 10\%.

\section{Conclusion}

At 500 GeV LC, it is crucial to instrument 
an active mask  to detect very forward electrons
down to 1\degr\ for measurement of
the small $\Delta M$.
The expected accuracy is 
10\% (20\% for $\Delta M \sim$ 5 GeV) 
with 500 \invfb.

\section*{Acknowledgments}
This work is supported in part by a NSF Grant
PHY-0101015, in part by  NSERC 
of Canada and in part by a DOE Grant DE-FG02-95ER40917.

\begin{table}
\caption{Accuracy of the $\Delta M$ determination 
for different beam mask designs. ``-'' means we cannot determine
with 500 \invfb.}
\label{table:dM_accuracy}
\begin{center}
\begin{tabular}{ | c| c| c c| }
\hline 
	& $N_{\stauonep\stauonem}$ &
	\multicolumn{2}{c|}{$\Delta M$(``500 \invfb'' expt.)} \\ \cline{3-4} 
 $\Delta M$ & (500 \invfb) & 2$^\circ$ mask  & 1$^\circ$ mask \\
\hline
 4.76  & 122 & - 	& 4.74$^{+0.97}_{-1.03}$\\
 9.53  & 787 & 9.5$^{+1.1}_{-1.0}$	& 9.5$^{+1.0}_{-1.0}$\\
 12.4 & 1027 & 12.5$^{+1.4}_{-1.4}$	& 12.5$^{+1.1}_{-1.4}$\\
 14.3 & 1138 & 14.5$^{+1.1}_{-1.4}$	& 14.5$^{+1.1}_{-1.4}$\\
\hline
\end{tabular}
\end{center}\end{table}


\begin{thebibliography}{99}
\bibitem{sp} D.N Spergel \etal, {\it Astr. J. Suppl.} {\bf 148}, 175 
(2003).
\bibitem{dark}
For example, see R. Arnowitt \etal, hep-ph/0308159.
\bibitem{sugra1} A.H. Chamseddine, R. Arnowitt, and P. Nath,
{\it Phys. Rev. Lett.} {\bf 49}, 970 (1982). 
\bibitem{sugra2} R.~Barbieri, S.~Ferrara, and C.A.~Savoy,
{\it Phys. Rev. D} {\bf 119}, 343 (1982); L. Hall, J. Lykken, and S. Weinberg,
{\it Phys. Rev. D} {\bf 27}, 2359 (1983); P. Nath, R. Arnowitt, and A.H. Chamseddine,
{\it Nucl. Phys. B} {\bf 227}, 121 {(1983)}.
\bibitem{higgs1} P.~Igo-Kemenes, LEPC meeting.
\bibitem{bsgamma} M. Alam \etal, {\it Phys. Rev. Lett.} {\bf 74}, 2885 {(1995)}. 

\bibitem{aleph} ALEPH collaboration, ALEPH-CONF 2001-009.
\bibitem{BNL} G. Bennett \etal, 
{\it Phys. Rev. Lett.} {\bf 92}, 161802 (2004).
\bibitem{isajet} F. Paige \etal, hep-ph/0312045.
We use version 7.69.
\bibitem{wphact}
E. Accomando, A. Ballestrero, and E. Maina, {\it Comput. Phys. Commun.} 
{\bf 150}, 166 (2003).
We use version 2.02pol.
\bibitem{TAUOLA}
        M. Je\.zabek \etal,  {\it Comput. Phys. Commun.}
        {\bf 76}, 361 (1993).  
\bibitem{JADE}
	JADE Collaboration, 
        {\it Phys. Lett. B} {\bf 213}, 235 (1988).

\end{thebibliography}
\end{document}